\documentclass[twocolumn,showpacs,preprintnumbers,superscriptaddress,
amsmath,amssymb]{revtex4}

\usepackage{epsfig}
\usepackage{graphicx}
\usepackage{dcolumn}
\usepackage{bm}
\usepackage{times}
\usepackage{xcolor}
\definecolor{light-gray}{gray}{0.8}

\begin{document}


\title{Social contagions on weighted networks}

\author{Yu-Xiao Zhu}
\affiliation{Web Sciences Center, University of Electronic Science
and Technology of China, Chengdu 611731, China}
\affiliation{Big Data Research Center, University of Electronic
Science and Technology of China, Chengdu 611731, China}

\author{Wei Wang}
\affiliation{Web Sciences Center, University of Electronic Science
and Technology of China, Chengdu 611731, China}
\affiliation{Big Data Research Center, University of Electronic
Science and Technology of China, Chengdu 611731, China}

\author{Ming Tang}
\email{tangminghan007@gmail.com}
\affiliation{Web Sciences Center, University of Electronic Science
and Technology of China, Chengdu 611731, China}
\affiliation{Big Data Research Center, University of Electronic
Science and Technology of China, Chengdu 611731, China}

\author{Yong-Yeol Ahn}
\email{yyahn@indiana.edu}
\affiliation{School of Informatics and Computing, Indiana University, Bloomington 47408, United States of America}

\date{\today}

\begin{abstract} We investigate critical behaviors of a social contagion model
on weighted networks. An edge-weight compartmental approach is applied to
analyze the weighted social contagion on strongly heterogenous networks with
skewed degree and weight distributions.  We find that degree heterogeneity can
not only alter the nature of contagion transition from discontinuous to
continuous but also can enhance or hamper the size of adoption, depending on
the unit transmission probability. We also show that, the heterogeneity of weight
distribution always hinder social contagions, and does not alter the
transition type.

\end{abstract}

\maketitle

\section{Introduction} \label{sec:intro}

Network provides a useful analytical framework for studying a wide array of
social phenomena, since the network of people---social networks---plays a
critical role in many social phenomena~\cite{newman_structure_2003,
Vespignani_RMP2015, CC_RMP2009, Boccaletti2006, Cohen2010,
barabasi-networkscience}. Although the edges in social networks---social
relationships---are often modeled binary, it is more realistic to
consider \emph{weighted} edges because the strength of social relationship
greatly vary in reality~\cite{Vespignani_pnas}. A number of proxies has been
used to capture the strength of social relationships. For example, the number
of papers that two scientists have coauthored was used to capture the strength
of the collaboration~\cite{Newman2001, Vespignani_pnas}; the duration of
calls---the amount of conversation---between two people is used to measure how
close they are~\cite{Onnela2007}. Thus it is important to ask how the
distribution of weights, along with degree distribution, affects various
dynamics on networks.

Spreading processes, such as epidemic spreading~\cite{Vespignani_RMP2015,
newman_network_2010, david_network_2006}, diffusion of
innovations~\cite{rogers_innovations,william_nature_1964,KRV}, and diffusion of
rumors~\cite{rumor_PRE,daley_nature_1964,rumor_epjb}, are fundamental dynamics
on social networks. Recent studies have shown that there exist two important
classes of contagions: simple and complex. Simple contagions (e.g.
epidemic models such as SIS model~\cite{Pastor-Satorras2001} and SIR
model~\cite{Moreno2002}) refers the processes where contagions spread
independently, while complex contagions (e.g. linear threshold
model~\cite{Granovetter_threshold,watts_threshold}) refers the processes that
are affected by \emph{social reinforcement}, where more exposures can
drastically increase the adoption probability~\cite{watts_threshold,
centola_science_2010,kleinberg_kdd_2006,KRV}.

Previous studies focused mainly on simple contagions, have revealed that
strong heterogeneity in the degree and weight distributions not only is
ubiquitous~\cite{Albert:2002,Vespignani_pnas,Vespignani_RMP2015}, but also
fundamentally affect the nature of spreading
phenomena~\cite{Pastor-Satorras2001,newman_structure_2003,wangwei_weight,
Gross2008, Holme2012}. For instance, on infinite scale-free networks where the
degree distribution exhibits a power-law ($P(k) \sim k^{-\alpha}, 2<\alpha\leq3
$), the epidemic threshold vanishes~\cite{Pastor-Satorras2001, Boguna2013}. The
inhomogeneity of weight distribution can also significantly affect the epidemic
threshold, epidemic prevalence, and spreading
velocity~\cite{wangwei_weight,Kamp2013,Rattana2013,Yan2005,Yang2012}.
Although many interesting properties of complex contagion has been uncovered
recently~\cite{centola_science_2010,gleeson_pre,weng_sr,azadeh}, it is not
fully understood how degree and weight heterogeneity affect the dynamics of
complex contagions.
Building on recent progress in complex
contagion~\cite{peterdodds_prl,Thomas_JMS1971,watts_threshold,KRV}, here
we introduce a weighted complex contagion model and investigate the effect of
degree and weight heterogeneity on the dynamics of complex contagion.

We find that (i) increasing heterogeneity of degree distribution changes the
nature of the phase transition from discontinuous to continuous; (ii) degree
heterogeneity plays opposite roles depends on the unit transmission probability:
it enhance the spreading when the unit transmission probability is small while
hinder the spreading when the unit transmission probability is large; and (iii)
the weight heterogeneity suppress the contagion while not altering the
transition type. To analyze the dynamics of complex contagion on weighted
networks, we use an edge-weight compartmental approach, which provides accurate
results.

\section{Weighted Complex Contagion Model and Network} \label{model}

We first introduce a complex contagion model that takes weighted edges into
account. Our model builds on a simple, generalized non-Markovian contagion
model that can describe both simple and complex
contagions~\cite{KRV,wangwei_NJP,wangwei_chaos}. In particular, an individual
can be in one of three possible states: \emph{susceptible} (S), \emph{adopted}
(A), or \emph{recovered} (R). Each individual has a state of awareness value
$m\in[0,T]$ which denotes the number of exposures. An individual adopts and
\emph{begins to transmit} the behavior or information (contagion) when its
awareness value reaches $T$. Individuals with $m<T$ do not affect the others.
Here we add a weight-based transmission rule---individuals transmit the
contagion preferably to its closer neighbors with the following probability:
\begin{equation} \lambda_{w_{i,j}} = 1-(1-\beta)^{w_{i,j}}, \end{equation}
where $w_{i,j}$ is the weight of the connection between individual $i$ and $j$,
and $\beta$ is the unit transmission probability. Given $\beta$,
$\lambda_{w_{i,j}}$ monotonically increases with $w_{i,j}$, i.e., individuals
are more likely to transmit the contagion to more strongly connected neighbors.
When successful, the awareness value of the neighbor will increase by one.
Assume an edge that has transmitted the contagion successfully will never
transmit the same information again. Also, each adopted
individual may become recovered with probability $\gamma$,
considering the fact that people may lose interest in the contagion
after a while and will not spread it any more (in this paper, we
set $\gamma=1$ unless noted, so everyone is active for only one step). The
individuals will remain in recovered state for all subsequent times once it is
recovered.

In our networks, we initially select a small fraction of nodes randomly and
designate them as \emph{seeds} by setting their awareness to be $T$. We set the
awareness of the remaining nodes to be $0$ and let them be at the susceptible
state. In each step, all adopted nodes will interact with all of its
susceptible neighbors and transmit the contagion to them with the probability
defined above. At the same time, all adopted nodes will recover with certain
probability. The spreading process stops when there is no adopted nodes, the
final adoption size is equal to final density of recovered nodes.

For simplicity, we assume uncorrelated random graphs specified by two
distributions: degree and weight. We realize such networks by generalizing
the configuration model~\cite{UCM, wangwei_weight}. Consider one network with
$N$ nodes and $M$ edges. We first create a graph using the classical
configuration model, where the degree distribution follows $p(k)\sim
k^{-\alpha_{k}}$ ($3\leq k_{i} \leq \sqrt{N}$),
then distribute weights that are sampled from $g(w)\sim w^{-\alpha_{w}}$
randomly ($w_{max} \sim N^{\frac{1}{\alpha_{w}-1}}$). $\alpha_{k}$
($\alpha_{w}$) controls the heterogeneity of the degree (weight) distribution.
Following previous studies~\cite{Yang2012,wangwei_weight}, we assume integer
weight values as it makes our approach more tractable.

\section{Theoretical Approach and Numerical Simulation}
\subsection{Edge-weight compartmental approach}
\label{sec:MF_theory}

One of the most widely used approaches to study network
dynamics---heterogeneous mean-field theory
(HMF)~\cite{Moreno2002,Castellano2006}---separates nodes into each degree
bucket while treating all edges equally. While it provides an excellent way to
handle strong degree heterogeneity, it overlooks edge weight heterogeneity. As
a result, the approach exhibits a limitation in dealing with networks with
strong weight heterogeneity~\cite{Buono2013}. Our edge-weight
compartmental approach treat each (integer) weight values separately and
provides a better way to study networks with strong weight
heterogeneity~\cite{Volz2008, Volz2011, wangwei_weight, wangwei_memory}.

We use variables $S(t)$, $A(t)$ and $R(t)$ to denote densities of the
susceptible, adopted, and recovered nodes at time $t$. Let us consider a
randomly selected susceptible node $u$ with awareness value $m$. Node $u$ will
remain susceptible as long as $m<T$ and will become adopted once $T$ of its
neighbors have transmitted the contagion successfully to $u$ since multiple
transmission through an edge is forbidden. As edge weights are assigned
randomly, the probability that $u$ is not informed by a neighbor $v$ by time
$t$ can be denoted by
\begin{equation} \label{theta}
\theta(t)=\sum_{w}g(w)\theta_{w}(t),
\end{equation}
where $\theta_{w}(t)$
denotes the probability that $u$ is not informed by an edge with weight $w$ by
time $t$. If $u$'s degree is $k$, the probability that the node was not one of
the seeds and received the contagion for $m$ times by time $t$ is
\begin{equation} \phi_{m}(k, t) =
(1-\rho_{0})\binom{k}{m}[\theta(t)]^{k-m}[1-\theta(t)]^{m},
\end{equation}
where $A_{0}$ denotes the fraction of seeds. Clearly, the probability that
the $k$-degree node was not one of the seeds and still didn't adopt the
contagion by time $t$ is
\begin{equation}
\phi(k, t) =\sum_{m=0}^{T-1}\phi_{m}(k, t).
\end{equation}
Thus the fraction of
susceptible nodes (the probability that a randomly selected node is
susceptible) at time $t$ is
\begin{equation} \label{susceptible}
S(t)=\sum_{k=0}p(k)\phi(k, t).
\end{equation}

Now, let us examine $\theta_{w}(t)$ in Eq.(\ref{theta}),
$\theta_{w}(t)$ can be broken down into:
\begin{equation} \label{sum_theta}
\theta_{w}(t)=\xi_w^{S}(t)+ \xi_{w}^{A}(t)+\xi_{w}^{R}(t),
\end{equation}
where $\xi_w^{X}(t)$ denote the probability that a neighbor in the state $X\in\{$S$,$A$,$R$\}$ has not
transmitted the contagion to $u$ through an edge with weight $w$ by time $t$.
Once we derive $\xi_w^{X}(t)$s, we can get the density of susceptible nodes at
time $t$ by substituting them into Eq.~(\ref{theta})-(\ref{susceptible}).

Neighbors who were in the susceptible state cannot inform $u$ unless they
themselves become adopted firstly. So first let us calculate the probability
that the neighbor remains to be susceptible by $t$. As we assume no correlation
between the degrees of nodes and its neighbors exists in uncorrelated networks, the
probability that a random neighbor of $u$ has degree $k$ is $kp(k)/\langle
k\rangle$, where $\langle k\rangle$ is the mean degree of the network. With
mean-field approximation, $\xi_w^{S}(t)$ is simply the probability that one of
its neighbors remains in the susceptible state by time $t$, which is given by
\begin{equation} \label{xiS}
\xi_w^{S}(t)=\frac{\sum_{k}kp(k)\phi(k-1, t)}{\langle k\rangle}.
\end{equation}
Note that, as we already know $u$ is in susceptible state at this time, so the probability that
this $k$-degree neighbor still didn't adopt the behavior by time $t$ is $\phi(k-1, t)$.

Calculating $\xi_{w}^{R}(t)$ requires considering two consecutive events:
first, an adopted neighbor has not transmitted the contagion to node $u$ via
their edge with weight $w$ with probability $1-\lambda_{w}$; second, the
adopted neighbor has been recovered, with probability $\gamma$. Combining
these two events, we have
\begin{equation} \label{xiR}
\frac{d\xi_{w}^{R}(t)}{dt}=\gamma[1-\lambda_{w}]\xi_{w}^{A}(t).
\end{equation}
If this adopted neighbor transmits the contagion via an edge with weight $w$,
the rate of flow from $\theta_w(t)$ to $1-\theta_{w}(t)$ will be
$\lambda(w)\xi_{w}^{A}(t)$, which means
\begin{equation} \label{theta_Rate}
\frac{d\theta_{w}(t)}{dt}=-\lambda_{w} \xi_{w}^{A}(t),
\end{equation}
and
\begin{equation} \label{theta_Rate1}
\frac{d(1-\theta_{w}(t))}{dt}=\lambda_{w}\xi_{w}^{A}(t).
\end{equation}
By combining Eqs.~(\ref{xiR})
and~(\ref{theta_Rate1}), one obtains
\begin{equation} \label{xiR_C}
\xi_{w}^{R}=\frac{\gamma[1-\theta_{w}(t)][1-\lambda_{w}]}{\lambda_{w}}.
\end{equation}
Substituting Eq.~(\ref{xiS}) and Eq.~(\ref{xiR_C}) into
Eq.~({\ref{sum_theta}}), we yield the following relation
\begin{equation}
\label{xiI} \xi_{w}^{A}(t)=\theta_w(t)- \frac{\sum_{k}kp(k)\phi(k-1,
t)}{\langle k\rangle}
-\frac{\gamma[1-\theta_{w}(t)][1-\lambda_{w}]}{\lambda_{w}}.
\end{equation}
By plugging this into Eq.~(\ref{theta_Rate}), we obtain
\begin{eqnarray}
\label{theta_w} \frac{d\theta_w(t)}{dt}&=&
\frac{\lambda_{w}\sum_{k}kp(k)\phi(k-1, t)}{\langle
k\rangle}-(1-\gamma)\lambda_{w}\theta_w(t)\nonumber \\
&+&\gamma[1-\lambda_{w}-\theta_w(t)].
\end{eqnarray}
From Eq.~(\ref{theta_w}),
the probability $\theta_w(t)$ can be computed.
The density associated with each distinct state is given by
\begin{equation}
\begin{cases}
\label{S_I_R}
\frac{dR(t)}{dt}  = \gamma A(t),\\
S(t)=\sum_{k=0}p(k)\phi(k,t),\\
A(t) = 1-R(t)-S(t).
\end{cases}
\end{equation}

From Eqs.~(\ref{theta_w}) and (\ref{S_I_R}), one can find that around $O(w_{max})$
equations are required in our edge-weight compartmental approach. By setting
$t\rightarrow \infty$ and $d\theta_w(t)/dt=0$ in Eq.~(\ref{theta_w}), we get
the probability of one edge with weight $w$ that didn't propagate the contagion in
the whole contagion process,
\begin{equation} \label{final_theta}
\theta_w(\infty)=\frac{\gamma[1-\lambda_{w}]+\frac{\lambda_{w}\sum_{k}kp(k)\phi(k-1,\infty)}{\langle
k\rangle}}{(1-\gamma)\lambda_{w}+\gamma}.\\
\end{equation}

$\theta_w(t)$ decreases with $t$ and thus if more than one stable fixed points
exist in Eq.~(\ref{final_theta}), only the maximum one is physically
meaningful~\cite{wangwei_memory,wangwei_NJP}.  Substituting $\theta_w(\infty)$
into Eqs.~(\ref{theta})-(\ref{susceptible}), we can calculate the value of
$S(\infty)$, and then final adoption size $R(\infty)$ can be obtained.
The number of roots in Eq.~(\ref{final_theta}) is either one or three.  If
Eq.~(\ref{final_theta}) has only one root, $R(\infty)$ increases continuously
with $\beta$, if Eq.~(\ref{final_theta}) has three roots, a saddle-node
bifurcation will occur, which leads to a discontinuous change in
$R(\infty)$\cite{theoretical_critical}.
The nontrivial solution corresponds to the point at which the equation
\begin{equation}
f(\theta(\infty))=\sum_{w}g(w){\frac{\gamma[1-\lambda_{w}]+\frac{\lambda_{w}\sum_{k}kp(k)\phi(k-1,\infty)}
{\langle k\rangle}}{(1-\gamma)\lambda_{w}+\gamma}}-\theta(\infty)
\end{equation} is tangent to horizontal axis at the critical value of
$\theta_c(\infty)$, in which $\theta_c(\infty)$ means the critical probability
that the information is not transmitted to $u$ via an edge at the critical
transmission probability when $t\rightarrow\infty$. We obtain the critical
condition of contagion by:
\begin{equation} \label{df}
\frac{df(\theta(\infty))}{d\theta(\infty)}|_{\theta_c(\infty)}=0.
\end{equation}
Plugging $\theta_c(\infty)$ into Eq.~(\ref{theta}) provides us with the
critical probability $\beta_{c}$.


\subsection{Simulation Results}

We report results of analytical solutions along with numerical simulations. We
consider networks with power-law degree and weight distributions: $p(k)\sim
k^{-\alpha_{k}}$ and $g(w)\sim w^{-\alpha_{w}}$. We use $A_{0} = 0.1$ across
paper but the results are robust with a range of $A_{0}$. Also we use, for each
parameter combination, $50$ network realizations, on each of which we run $100$
independent simulations.

\begin{figure}
\begin{center}
\includegraphics[trim=35mm 105mm 30mm 100mm, width=0.53\textwidth, height=0.32\textwidth]{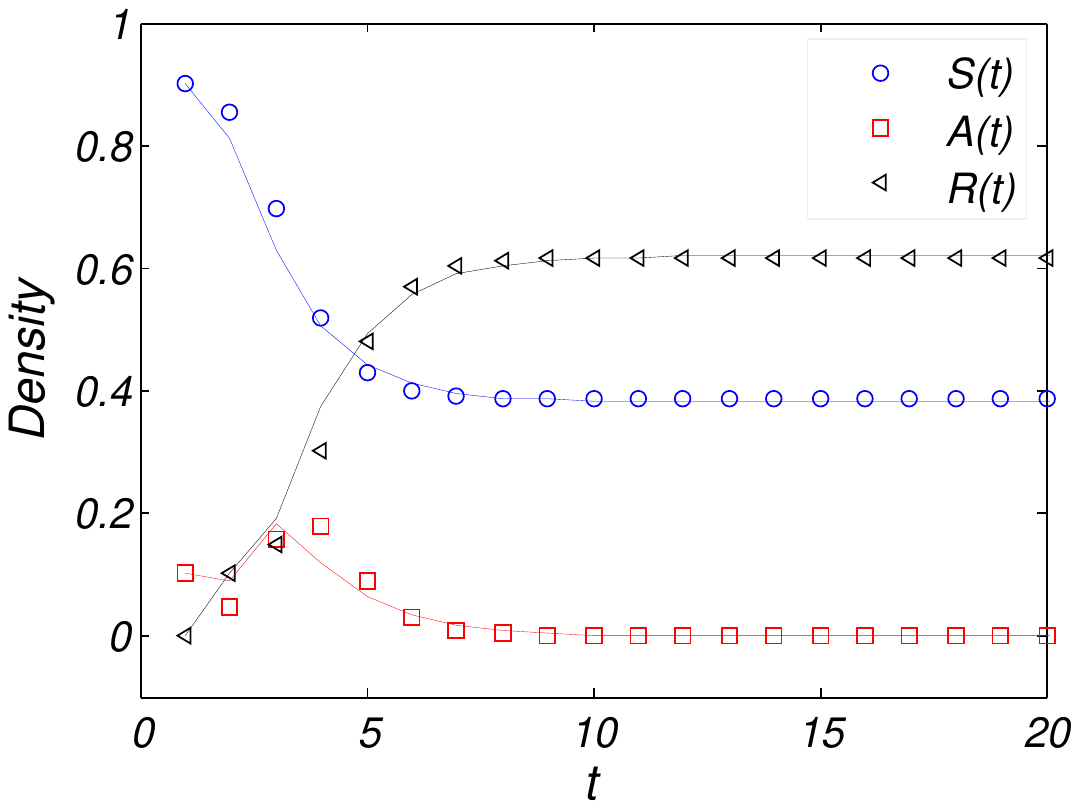}
\caption{Time evolutions of densities of nodes in different states, denoted by $S(t)$ (blue circles), $A(t)$ (red squares), and $R(t)$ (black left triangles), respectively. Analytical results are plotted in lines, which match well with simulation results (symbols).
The parameters for the simulations are $N$=10,000, $\langle k \rangle$ = 10, $\langle w \rangle$ = 8, $\alpha_{k}$=2.1, $\alpha_{w}$=2.4, $\beta$=0.18, $A_{0}$=0.1 and $T$=3.}
\label{t_SIR}
\end{center}
\end{figure}

\begin{figure}
\begin{center}
\includegraphics[trim=35mm 70mm 40mm 69mm, width=0.38\textwidth, height=0.48\textwidth]{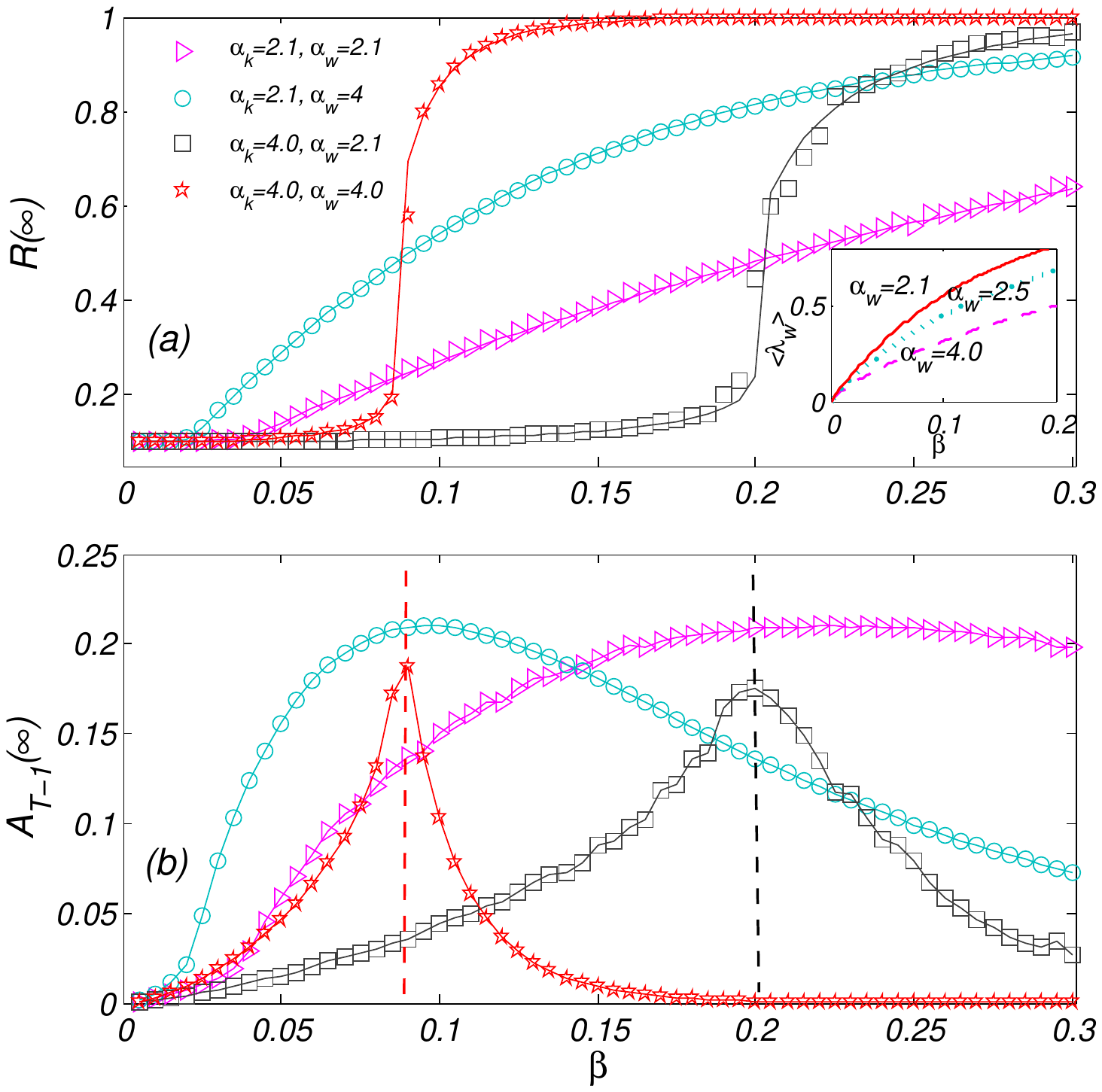}

\caption{Final states of complex contagion dynamics on weighted networks.
(a) Final adoption size ($R(\infty)$) versus the unit transmission probability
($\beta$) on different networks with tunable parameters.  The inset shows the
numerical solutions of mean transmission rate $\langle \lambda_w \rangle$ as
function of $\beta$ for three different values of $\alpha_{w}$ (i.e., 2.1, 2.5,
and 4.0). (b) Final subcritical size ($A_{T-1}(\infty)$) versus the $\beta$ on
different networks with tunable parameters. The parameters for all the
simulations are $N$=10,000, $\langle k \rangle$ = 10, $\langle w
\rangle$ = 8, $A_{0}$=0.1 and $T$=3.} \label{number_21_4}

\end{center}
\end{figure}
%

Figure~\ref{t_SIR} illustrates the time evolutions of susceptible (S), adopted
(A) and recovered (R) nodes. Naturally, it displays a very similar dynamics
with SIR model.  Our analytical results (lines) agree well with simulation
results (symbols).  In Fig.~\ref{number_21_4}(a), we show the final adoption
size ($R(\infty)$) in relationship with unit transmission probability ($\beta$)
for networks with different degree and weight distributions along with the
analytical results (shown in black line, which match well with the simulation
results).  Now, let us focus on the influence of heterogeneous degree
distribution on social contagion processes from two perspectives: the
transition type of $R(\infty)$ with $\beta$ and the final adoption size.  We
summarise our results as following.

\begin{figure}
\begin{center}
\includegraphics[trim=35mm 102mm 40mm 85mm, width=0.48\textwidth, height=0.35\textwidth]{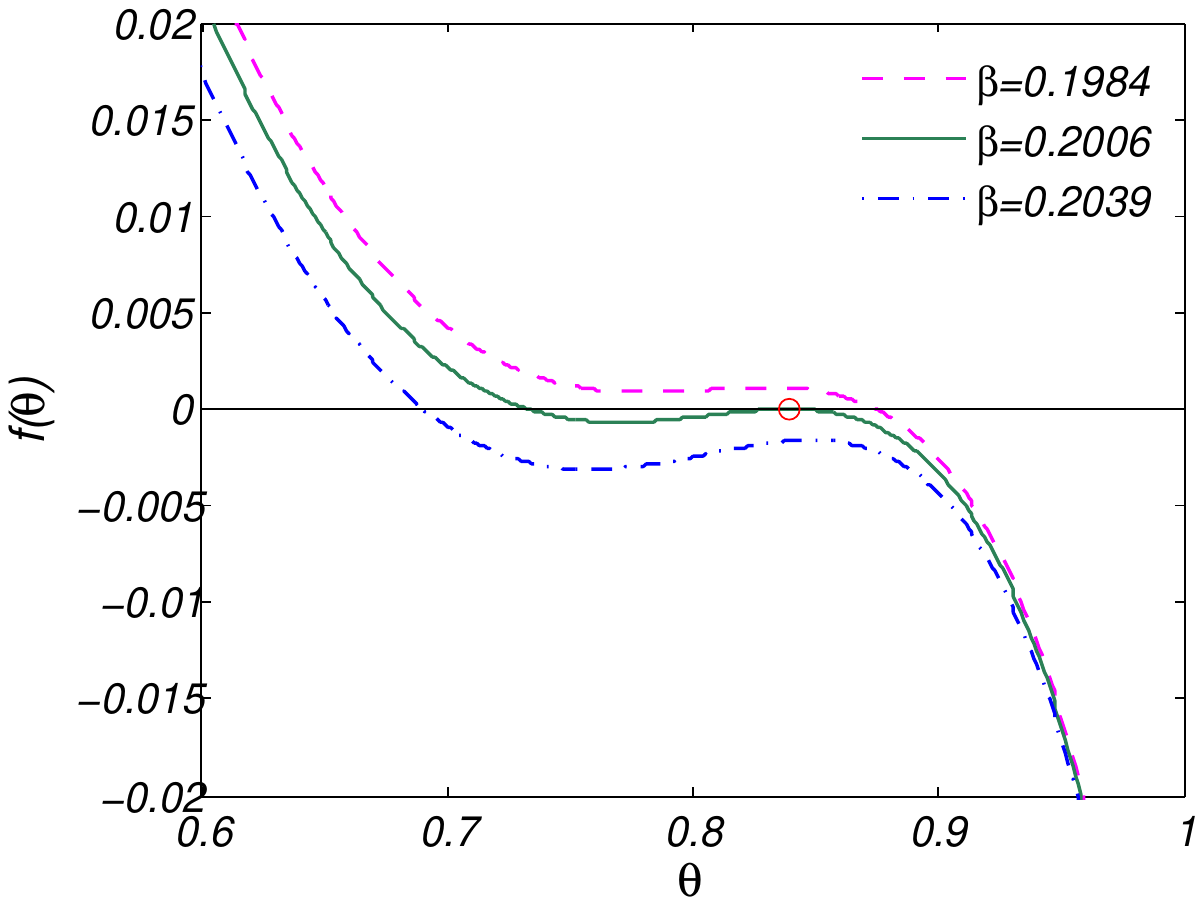}
\caption{Illustration of graphical solutions of Eq.~(\ref{final_theta}).
The black solid line is the horizontal axis and
the red circle denotes the tangent point. The parameters for the simulations are $N$=10,000, $\langle k \rangle$ = 10, $\langle w \rangle$ = 8,
$\alpha_{k}=4.0$, $\alpha_{w}=2.1$, $A_{0}$=0.1 and $T$=3.}
\label{critical_example}
\end{center}
\end{figure}

First, the degree exponent determine the discontinuity of the transition as
shown in a previous work~\cite{wangwei_memory}.  Figure~\ref{number_21_4}(a)
shows that $R(\infty)$ increases continuously with $\beta$ with heterogeneous
degree distribution (e.g., $\alpha_{k}=2.1$), while exhibiting a discontinuous
transition when $\alpha_{k}=4.0$.  The results of bifurcation analysis on
Eq.~(\ref{final_theta}) show that there exists one critical degree exponent
$\alpha_{k}^{c}\approx4.0$, below (above) which $R(\infty)$ versus $\beta$ is
continuous (discontinuous).  For networks with $\alpha_{k}=4.0$, the
value of $\beta_{c}$ can be obtained from Eq.~(\ref{final_theta}) using
bifurcation theory~\cite{theoretical_critical}.  Analytical calculations show
that for Eq.~(\ref{final_theta}), the number of roots in
Eq.~(\ref{final_theta}) is either one or three (see
Fig.~\ref{critical_example}).  If Eq.~(\ref{final_theta}) has only one root,
$R(\infty)$ increases continuously with $\beta$; if
Eq.~(\ref{final_theta}) has three roots, a
saddle-node bifurcation occurs\cite{theoretical_critical}.  As
shown in Fig.~\ref{critical_example}, there is only one fixed point of
Eq.~(\ref{final_theta}) at a small value of
$\beta$ (e.g, $\beta=0.1984$) and then three fixed
points (in this case, only the maximum one is physically meaningful since
$\theta(t)$ decreases with $t$) gradually emerge with the increasing of
$\beta$.  The tangent point that marked as one red circle is
the physically meaningful solution at the
unit transmission probability $\beta_c$ (e.g,
$\beta=0.2006$).  For $\beta> \beta_c$ (e.g, $\beta=0.2039$), the solution of
Eq.~(\ref{final_theta}) changes to a smaller solution abruptly, which leads to
a discontinuous change in $R(\infty)$.  We can demonstrate the type of
dependence and obtain the value of $\beta_c$ for other
parameters through the similar measure.

\begin{figure}
\begin{center}
\includegraphics[trim=40mm 103mm 48mm 80mm, width=0.32\textwidth, height=0.32\textwidth]{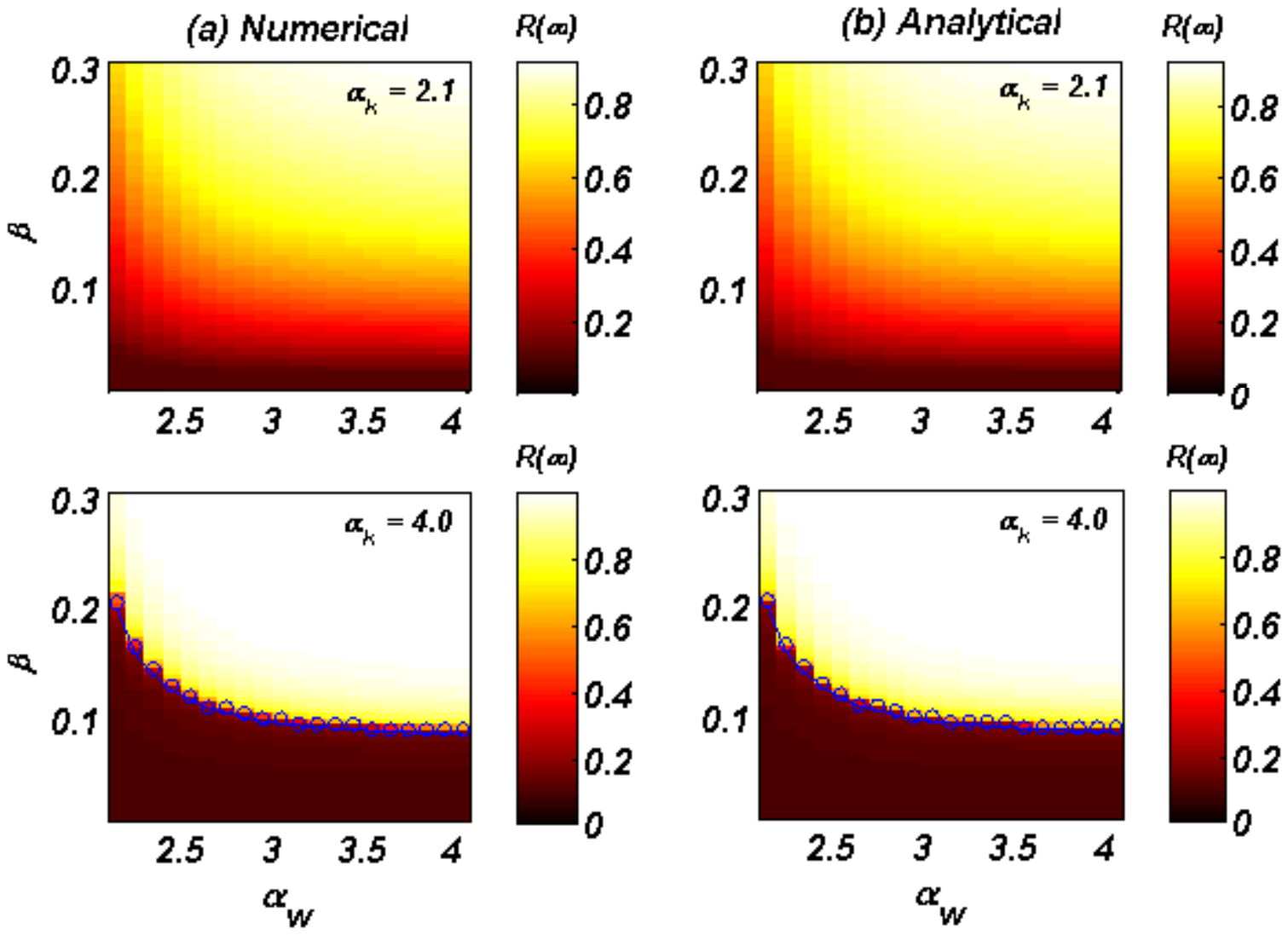}

\caption{ The relationships between $\alpha_{w}$, $\beta$ and $R(\infty)$ at a
fixed $\alpha_{k}$. Figures (a) and (b) are results of numerical simulation
and analytical method on networks with two different $\alpha_{k}$ (i.e., 2.1
and 4.0) respectively. The parameters for simulation are $N$=10,000,
$\langle k \rangle$ = 10, $\langle w \rangle$ = 8, $A_{0}$=0.1 and $T$=3.}
\label{heatmap_w_p}

\end{center}
\end{figure}

We also explain this phenomena visually by showing $A_{T-1}(\infty)$ (final
subcritical size) in Fig.~\ref{number_21_4}(b).  Here nodes with awareness
value $T-1$ are considered as in subcritical state.  Clearly, for results of
$\alpha_{k}=4.0$ in Fig.~\ref{number_21_4}(b), the quick sharp decline of final
subcritical size corresponds to a dramatic increase of final adoption size,
thus may induce a discontinuous dependence.  Note that, there is no so-called
critical value of unit transmission probability ($\beta_c$) for continuous
dependence. The critical value $\beta_{c}$ can also be estimated by increasing
the number of iterations~\cite{havlin2010} (only those interactions in which
appears at least one newly adopted individual are taken into account). In
Fig.~\ref{number_21_4}(b), we show the estimated $\beta_{c}$ with dashed lines,
which correspond to the peaks of $A_{T-1}(\infty)$.  More details are shown in
Fig.~\ref{heatmap_w_p}.  As we expected, $R(\infty)$ for networks with
heterogeneous degree distribution (Fig.~\ref{heatmap_w_p}(a) above) shows a
continuous change with the increasing of $\beta$ while change discontinuously
on networks with homogeneous degree distribution (Fig.~\ref{heatmap_w_p}(a)
below).  Analytical results shown in Fig.~\ref{heatmap_w_p}(b) agree well with
numerical results.  The estimated values of $\beta_{c}$ are labeled in blue
circle, along with the corresponding analytical critical results are plotted in
blue line (shown in Figs.~\ref{heatmap_w_p}(b)).

\begin{figure}
\begin{center}
\includegraphics[trim=57mm 95mm 48mm 76mm, width=0.298\textwidth, height=0.32\textwidth]{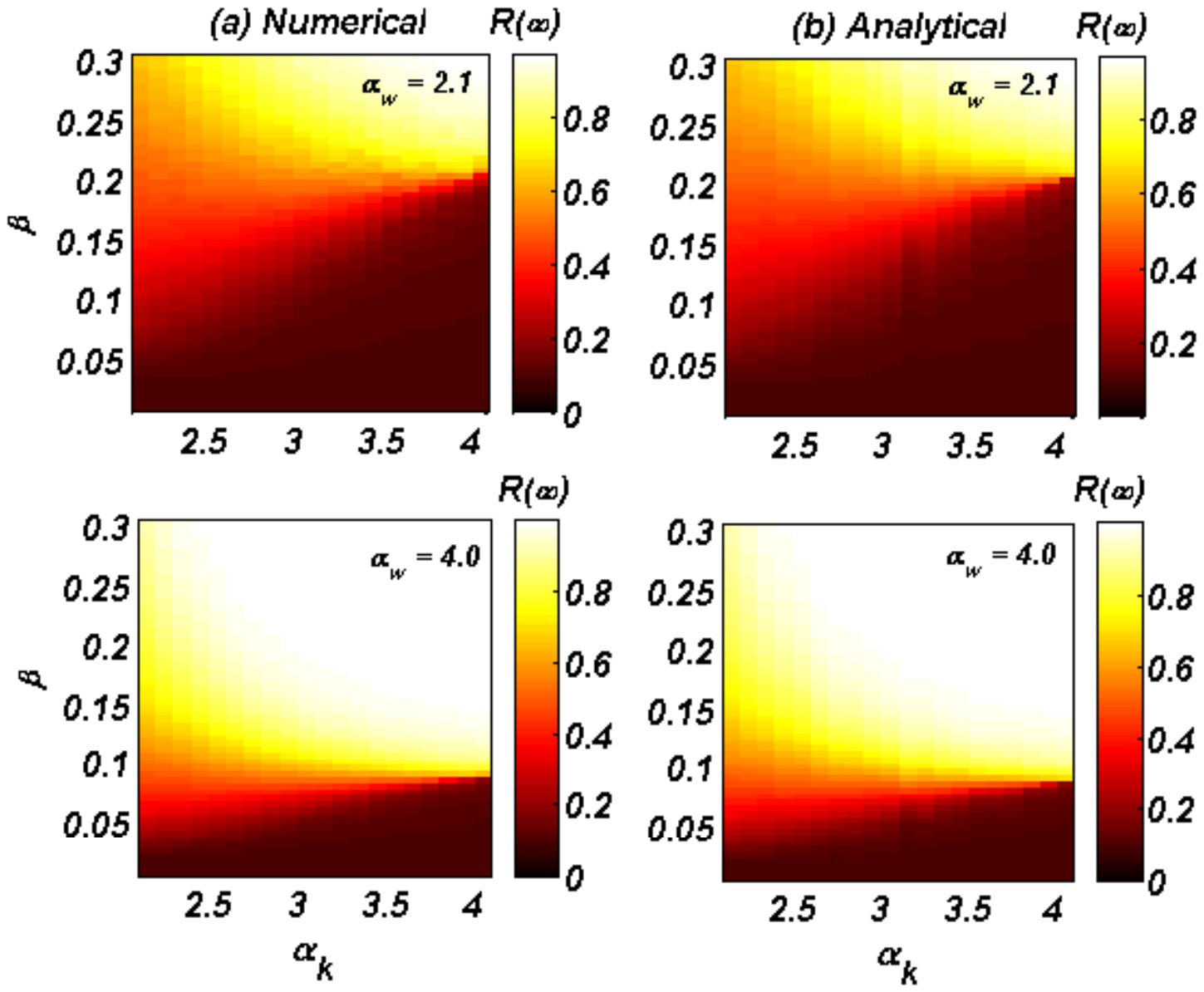}
\caption{The relationships between $\alpha_{k}$, $\beta$ and $R(\infty)$ at a
fixed $\alpha_{w}$. Figures (a) and (b) are results of numerical simulation
and analytical method on networks with two different $\alpha_{w}$ (i.e., 2.1
and 4.0), respectively. The parameters for the simulations are
$N$=10,000, $\langle k \rangle$ = 10, $\langle w \rangle$ = 8, $A_{0}$=0.1 and
$T$=3. } \label{heatmap_k_p}
\end{center}
\end{figure}

Second, as a result of continuous-discontinuous transition, degree
heterogeneity enhances the final adoption size at small $\beta$ while hindering
it at large $\beta$, which is consistent with that of epidemic
case~\cite{wangwei_weight}.  For instance, when $\alpha_{w}=2.1$, the final
adoption size ($R(\infty)$) for $\alpha_{k} = 2.1$ is greater than that of
$\alpha_{k} = 4.0$ when $\beta < 0.2$, while opposite situation is obtained
when $\beta > 0.2$ (shown in Fig.~\ref{number_21_4}(a)).  This result can be
qualitatively explained as following: social contagion propagates on complex
networks in two-stages due to the co-emergence of more hubs and large amount of
small-degree nodes with increasing heterogeneity of degree distribution. The
hubs are more likely to become adopted early since more neighbors make them
have higher chance to reach the identical awareness threshold $T$ thus get
adopted. On the contrary, small-degree nodes are less likely to become adopted
due to the small number of its neighbors. Given a network with heterogeneous
degree distribution, when unit transmission probability $\beta$ is small, the
existence of more hubs enhance the contagion thus leads to greater $R(\infty)$
(promotion region); When $\beta$ is large, the existence of large amount of
small-degree nodes will hinder the contagion, resulting in smaller $R(\infty)$
(suppression region).  Fig.~\ref{heatmap_k_p} shows the whole picture of
relationship between $\alpha_{k}$, $\beta$ and $R(\infty)$ when fixing
$\alpha_{w}$.  Increasing the heterogeneity of degree distribution will enhance
$R(\infty)$ at small $\beta$ while hinder the adoption size at large $\beta$.

Let us address the influence of the heterogeneity of
weight distribution on social contagion processes at a given
$\alpha_{k}$. The heterogeneity of weight distribution (smaller value of
$\alpha_{w}$) reduces final adoption size $R(\infty)$. For instance, if
we fix $\alpha_{k}$ (Fig.~\ref{number_21_4}(a)), $R(\infty)$ for $\alpha_{w} =
2.1$ is always smaller than that of $\alpha_{w} = 4.0$. This phenomenon can be
explained as follows: when the average weight $\langle w
\rangle$ is fixed, in the network with smaller $\alpha_{w}$, most edges have
lower weights and thus transmission probabilities, leading to a smaller
mean transmission rate $\langle \lambda_{w} \rangle=\sum_{w}g(w)\lambda_{w}$
for a randomly selected edge. As shown in the inset of
Fig.~\ref{number_21_4}(a), $\langle \lambda_{w} \rangle$ of $\alpha_{w}$=2.1 is
smaller than that of $\alpha_{w}=4.0$ with a given $\beta$. On the other hand,
changing the weight distribution will not change the dependence behavior of
($R(\infty)$, $\beta$) with a given degree distribution, which is similar to
the case of simple contagion models~\cite{wangwei_weight}. This finding
can be verified from analytical perspective, varying the value of $\alpha_{w}$
will not change the number of roots in Eq.~(\ref{final_theta}), thus will not
affect whether saddle-node bifurcation occur or not. The relationship between
$\alpha_{w}$ and $\beta$ when fixing $\alpha_{k}$ is shown in
Fig~\ref{heatmap_w_p}, which confirms our finding here.

\begin{figure}
\begin{center}
\includegraphics[trim=35mm 90mm 30mm 76mm, width=0.38\textwidth, height=0.32\textwidth]{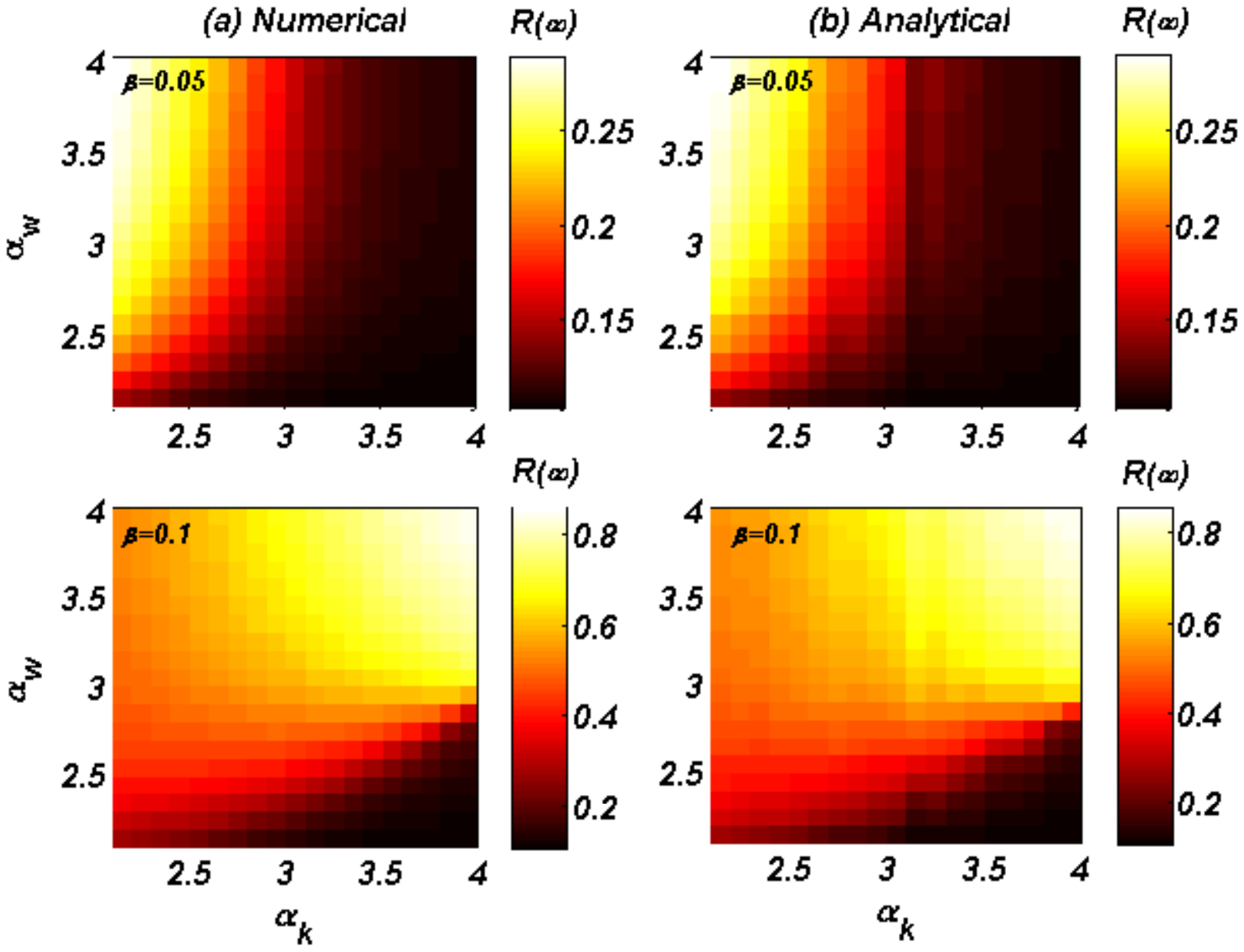}
\caption{The relationships between $\alpha_{k}$, $\alpha_{w}$ and $R(\infty)$ when the unit transmission probability $\beta$ is fixed. Figures (a) and (b) are results of numerical simulation and analytical method on the weighted
contagion model with two different $\beta$ (i.e., 0.05 and 0.1), respectively.
The parameters for the simulations are $N$=10,000, $\langle k \rangle$ = 10, $\langle w \rangle$ = 8, $A_{0}$=0.1 and $T$=3.}
\label{heatmap_k_w}
\end{center}
\end{figure}
Finally, Fig.~\ref{heatmap_k_w} summarizes
our results, showing that for small value of transmission probability ($\beta=0.05$),
the existence of more hubs that can be easily informed thus enhance
the contagion process (Promotion Region). While for large value of transmission probability ($\beta=0.1$),
the existence of large-amount small-degree nodes that difficult to be adopted will hinder
the contagion process (Suppression Region). In addition, increasing the heterogeneity of weight distribution will always hinder $R(\infty)$.
Figs.~\ref{heatmap_k_w}(a) and \ref{heatmap_k_w}(b) show results of simulation and analytical method respectively, which match well with
each other.

\section{Conclusions} \label{sec:conclusion}

In summary, we study the effect of
heterogenous network structures on the diffusion of complex contagions.
With decreasing heterogeneity of degree distribution,
the dependence of final adoption size on unit transmission probability changes from being continuous to discontinuous.
We then show that the heterogeneity of degree distribution
may have two opposite effects depending on the transmission probability:
degree heterogeneity enhances complex contagions when $\beta$ is small while
hindering it when $\beta$ is large.
By contrast, the heterogeneity of weight distribution always
reduces final adoption size though not change the dependence pattern of final adoption size on unit transmission probability.


Our findings offer insights to understand the influence of underlying network structures for weighed social contagions.
Future work may investigate into the cases where
the adoption threshold of each individual varies with its degree, or a
richer and correlated network structure is assumed.

\acknowledgments

This work was partially supported by the National Natural Science Foundation of China
(Grant Nos. 11105025, 11575041 and 61433014) and the Program of Outstanding PhD Candidate
in Academic Research by UESTC: YBXSZC20131035. YYA thanks Microsoft Research for
MSR Faculty Fellowship.

\bibliographystyle{apsrev4-1}
\bibliography{Spreading}

\end{document}